# A Panel Prototype for the Mu2e Straw Tube Tracker at Fermilab

*Alessandra Lucà (\*)*

*Fermi National Accelerator Laboratory, Batavia, Illinois 60510, USA.*
*(\*) aluca@fnal.gov*

## 1. INTRODUCTION

The Mu2e experiment will search for coherent, neutrino-less conversion of muons into electrons in the Coulomb field of an aluminum nucleus with a sensitivity of four orders of magnitude better than previous experiments. The signature of this process is an electron with energy nearly equal to the muon mass. Mu2e relies on a precision (0.1%) measurement of the outgoing electron momentum to separate signal from background. In order to achieve this goal, Mu2e has chosen a very low-mass straw tracker, made of 20,736 5 mm diameter thin-walled (15 μm) Mylar® straws, held under tension to avoid the need for supports within the active volume, and arranged in an approximately 3 m long by 0.7 m radius cylinder, operated in vacuum and a 1 T magnetic field. Groups of 96 straws are assembled into modules, called panels. We present the prototype and the assembly procedure for a Mu2e tracker panel built at Fermilab.

## 2. GENERAL FEATURES OF THE TRACKER DESIGN

The selected design for the Mu2e tracker is a low mass array of straw drift tubes [1]. The drift gas is taken as 80:20 Argon:$CO_2$ with an operating voltage of ≤1500 V.

The basic tracker element is a 25 μm gold plated tungsten sense wire centered in a 5 mm diameter tube, referred to as a *straw*. Each straw is made of two layers of ~6 μm (25 gauge) Mylar®, spiral wound, with a ~3 μm layer of adhesive between layers. The total thickness of the straw wall is 15 μm. The inner surface has 500 Å of aluminum overlaid with 200 Å of gold as the cathode layer. The outer surface has 500 Å of aluminum to act as additional electrostatic shielding and to reduce the leak rate. The straws vary in active length from 334 mm to 1174 mm and are supported only at the ends.

Groups of 96 straws are assembled into roughly trapezoidal panels. Each panel covers a 120° arc and has two layers of straws to improve efficiency and help determine on which side of the sense wire a track passes (the classic *left-right ambiguity*). A 1 mm gap is maintained between straws to allow for manufacturing tolerances and expansion due to gas pressure. This necessitates that individual straws be self-supporting across their span.

Figure 1 and Table 1 show the breakdown of components in the tracker. The tracker consists of 18 *stations*, evenly spaced along its whole length of ~3 m, and associated infrastructure. Each station is made of two *planes* (36 planes total) and a plane consists of 6 *panels* (216 panels total) rotated by 30°, on two faces of a support ring; there are three panels per face. The total number of straws is 20,736.

Table I: Breakdown of the number of components in the Mu2e tracker. The straw total of 20,736 at the bottom of the second column is the product of the numbers above.

|  |  |
| --- | --- |
| Stations | 18 |
| Planes per station | ×2 |
| Panels per plane | ×6 |
| Straws per panel | ×96 |
| **Total straws** | **20,736** |

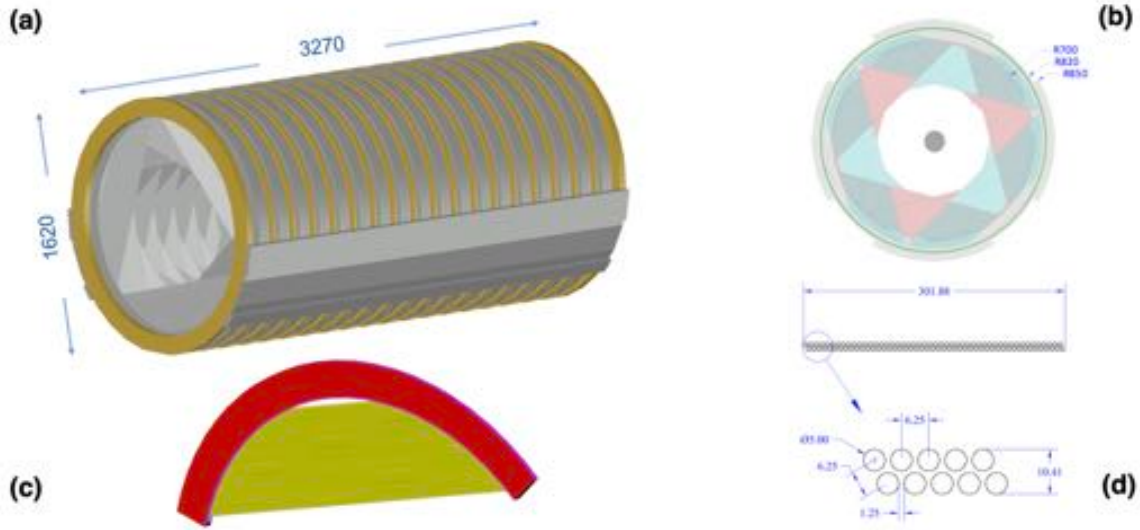

Figure 1: The CAD drawing of the assembled Mu2e tracker, with 18 stations (a). A single station is made of 12 panels with different rotation angles (b). A panel (c), with cover in red and straws in yellow, has 96 straws arranged as shown in (d). Dimensions are in millimeters.

The tracker is designed to operate an overall mean time to failure (MTTF) of >1 year [2]. Moreover, it must be mechanically capable of withstanding a pressure differential of 1 ATM, and have a low enough leak rate so that leaks will not drive the pressure in the detector solenoid above $\geq 10^{-4}$ Torr[1].

---

[1] The estimated allowed leak rate, inclusive of virtual leaks and slow outgassing, is 6 cm$^3$/minute (or 0.03 cm$^3$/minute per panel).

## 3. ASSEMBLY OF PANEL V2.5

A tracker panel prototype has been built at Fermilab: panel v2.5. The assembly procedure, summarized below, has been performed using a dedicated aluminum plate called *Panel Assembly and Alignment System* (PAAS). The PAAS has various features to align panel parts and external fixtures, and ensures a correct positioning of straws and sense wires.

### 3.1. Inner Ring

For the inner ring assembly, first the *bottom inner ring*, which is a stainless steel arc, is attached to the PAAS. Then, plastic *manifold insert*s are aligned, and epoxied on top of it, together with a stainless steel filler, placed in the gap between plastic inserts, and referred as *middle inner ring*. The manifold inserts are produced with a 3D printer. As shown in Figure 2(a), their position is set connecting them to dedicated alignment devices (*vectra guides*), using dowel pins.

After applying the epoxy, the curing is performed via a controlled heating process: the PAAS and inner ring temperatures are monitored with suitable probes (i.e., thermocouples and infrared sensors) and gradually risen up to a steady state.

### 3.2. Straws

When the inner ring is set, the straws are pulled into the manifold holes, starting from the bottom layer. The manifold plastic inserts serve to position the straws and have in total 192 holes. Each straw has a 5 mm outer diameter brass tube inserted in each end. Inside the brass tube there is a longer plastic tube to deaden ~1 cm past the gas manifold, and to protect against breakdown. A grooved support plate keeps straws in place while pulled in.

The straws are positioned via external alignment tools: the so-called *combs*. The combs have a set of steel dowel pins. If the inner ring is out of tolerance, combs move straws close to the nominal position. Therefore, the location of the manifold holes is not critical; however, to avoid weakening the plate, a 0.25 mm placement and size accuracy is required.

After pulling and aligning the straws, they are held under at tension by external fixtures, the *anchor walls*. These devices are produced with selective laser sintering and consist on a slots system. The straw terminations are latched to the ends of the corresponding slot and set to a fixed-distance with respect the slot end. Figure 2(b) shows the straws alignment and tensioning stage.

When under tension, straws are epoxied to the manifold inserts holes, with the epoxy also forming the gas seal.

### 3.3. Sense wires

To avoid accumulating errors, straws and sense wires are aligned separately: the sense wire location is set by the combs, and not by the straw termination.

A 4.95 mm outer diameter brass tube is mechanically and electrically connected to each straw end using silver epoxy. Inside the brass tube there is an extruded Kapton® tube to protect against breakdown at the edge of the brass tube, and inside the Kapton® tube there is an injection molded plastic insert. A small U-shaped brass pin is attached to a groove in the insert.

The sense wire is pulled inside the straw and soldered to the pin, after the straws installation is completed. For this process, the wire is tensioned to 80 g and supported only at the ends. The brass pin allows up to 250 μm mispositioning of the straw body without impacting the sense wire location.

### 3.4. Gas manifold and preliminary tests

Figure 2(c) shows the gas manifold finishing. After the wires installation is completed, the baseplate is positioned, and glued the to the inner ring; then, the frame is glued to the baseplate and inner ring. The frame is made from a single piece of aluminum and provides a smooth, uniform surface for the O–ring groove. Finally, O–rings and covers complete the manifold gas seal. Covers are bolted down with screws.

Preliminary tests of the completely assembled panel 2.5 have been performed. Placing the prototype in a vacuum chamber, it demonstrates to meet the design requirements: it is mechanically capable of withstanding a pressure differential of 1 ATM and the leak rate, inclusive of virtual leaks and slow outgassing, does not exceed 0.03 $cm^3$/minute per panel.

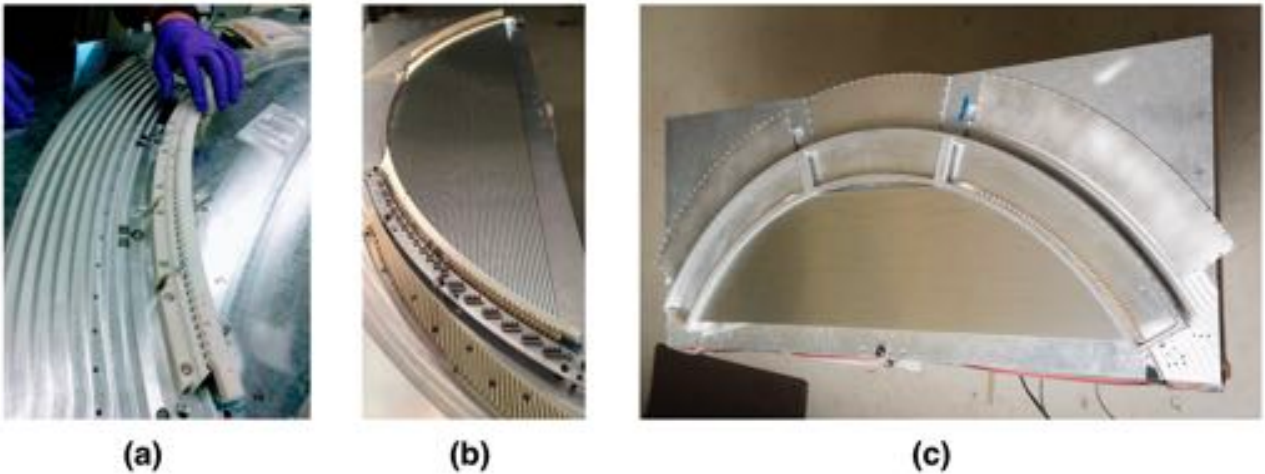

Figure 2: Some steps of the assembly procedure of the tracker panel prototype built at Fermilab - panel 2.5 - : inner ring assembly (a), straws installation (b), and gas manifold finishing (c) with covers not yet in place.

### 4. CONCLUSION

A prototype for the Mu2e tracker, panel v2.5, has been built at Fermilab. The mechanical assembly fulfills the requirements and is intended to be the final design.

**Ackowledgments**

We are grateful for the vital contributions of the Fermilab staff and the technical staff of the participating institutions. This work was supported by the US Department of Energy; the Italian Istituto Nazionale di Fisica Nucleare; the Science and Technology Facilities Council, UK; the Ministry of Education and Science of the Russian Federations; the US Science Foundation; the Thousand Talents Plan of the Republic of China; the Helmholtz Association of Germany; and the EU Horizon 2020 Research and Innovation Program under the Marie Sklodowska-Curie Grant Agreement No.690385. Fermilab is operated by Fermi Research Alliance, LLC under Contract No. DE-AC02-07CH11359 with the US Department of Energy.